\documentclass[aps,prc,tightenlines,floats,floatfix,showpacs]{revtex4}
\usepackage{amsmath}
\usepackage{bm}
\usepackage{dcolumn}
\usepackage{psfig}
\usepackage{graphicx}

\renewcommand{\Re}{{\rm{Re}}}

\begin{document}
\preprint{\Large{Internal report to e91011 collaboration}}

\title{Accuracy of traditional Legendre estimators of quadrupole ratios
for the $N \rightarrow \Delta$ transition}
\author{J. J. Kelly}
\affiliation{Department of Physics, University of Maryland, College Park,
Maryland 20742, USA}
\date{July 6, 2005}

\begin{abstract}
We evaluate the accuracy of traditional estimators often used to extract
$N \rightarrow \Delta$ quadrupole ratios from cross section angular 
distributions for pion electroproduction.
We find that neither $M_{1+}$ dominance nor $\ell \leq 1$ truncation is 
sufficiently accurate for this purpose.
Truncation errors are especially large for $R_{EM}$, for which it
is also essential to perform Rosenbluth separation.
The accuracy of similar truncated Legendre analyses for $E_{0+}$, $S_{0+}$, 
and especially $M_{1-}$ is even worse.
\end{abstract}

\maketitle

Historically the most important indications of deformation of low-lying
baryons have been the quadrupole ratios for electromagnetic excitation of
the $N \rightarrow \Delta(1232)$ transition.
Magnetic dipole excitation dominates and is represented by the $M_{1+}$ 
multipole amplitude while nonzero values for the electric and scalar 
(longitudinal) multipoles, $E_{1+}$ and $S_{1+}$, arise either from 
nonspherical contributions to the wave functions or from higher-order 
dynamical contributions to the electromagnetic transition.
The quadrupole ratios are defined as
\begin{subequations}
\begin{eqnarray}
R_{EM} = \Re \frac{E_{1+}}{M_{1+} } \\
R_{SM} = \Re \frac{S_{1+}}{M_{1+} }
\end{eqnarray}
\end{subequations}
evaluated at the physical mass of the resonance, 
$W=M_\Delta\approx 1.232$ GeV.
Determination of quadrupole ratios for isospin-3/2 amplitudes requires
measurements of two charge states, such as $p\pi^0$ and $n\pi^+$. 
Complex multipole amplitudes have been deduced for $Q^2 = 0$ using
polarization data for pion photoproduction \cite{Blanpied01},
but few experiments for $Q^2 > 0$ have provided sufficient information
to perform an actual multipole analysis.
Instead, most experimental determinations of $N \rightarrow \Delta$
transition form factors 
\cite{Kalleicher97,Frolov99,Joo02,Sparveris05} rely upon estimators 
derived from multipole expansions for the angular dependence of unpolarized 
cross sections using two simplifying assumptions: 
1) only multipoles with $\ell \leq 1$ contribute, which is described as 
$sp$ truncation; and
2) only terms involving $M_{1+}$ are retained, which is described as
$M_{1+}$ dominance.
Although the reliability of these assumptions has been questioned before,
the improved kinematic completeness and statistical precision of modern
experiments warrants re-examination of their accuracy.
In this Brief Report, we consider the accuracy of traditional quadrupole
estimators for $Q^2 \sim 1$ (GeV/$c$)$^2$ where a nearly model-independent 
multipole analysis of recoil-polarization response functions for the 
$p(\vec{e},e^\prime \vec{p})\pi^0$ reaction disagrees appreciably with the 
traditional Legendre analysis of the cross section data 
\cite{Kelly05c,Kelly05d}.
 
The unpolarized differential cross section for $\gamma_v N \rightarrow N\pi$ 
in the $\pi N$ center of momentum frame takes the form
\begin{equation}
\frac{d\sigma}{d\Omega_\pi} = \nu_0 \left(
\epsilon_S R_L + R_T +
\sqrt{2\epsilon_S (1+\epsilon)}R_{LT} \sin{\theta}\cos{\phi}
+ \epsilon R_{TT} \sin^2{\theta}\cos{2\phi} \right)
\end{equation}
where $\nu_0$ is a phase-space factor, $\epsilon$ is the transverse
polarization of the virtual photon, $\epsilon_S = \epsilon Q^2 / q^2$, 
and $(\theta,\phi)$ are polar and azimuthal pion angles relative to
the $\vec{q}$ vector and the electron scattering plane.
The response functions can be expanded in Legendre series
\begin{equation}
R_\lambda = \sum_{n=0}^{\infty} A^\lambda_n P_n(\cos{\theta})
\end{equation}
where $\lambda \in \{ L,T,LT,TT \}$.
The expansion coefficients, $A^\lambda_n$, are functions of $(W,Q^2)$ that
can be fit to the angular distribution of the differential cross section.
Each of those coefficients can in turn be expressed as a multipole 
expansion containing terms of the form $\Re B_{\ell\pm}C_{\ell\pm}^*$ 
where $B,C \in \{M,E,S\}$ are magnetic, electric, or scalar multipole 
amplitudes for specified $\ell$ and $j=\ell \pm 1/2$.
In principle, experimentally determined Legendre coefficients include 
contributions from arbitrarily large $\ell$ and are not limited either 
by $sp$ truncation or by $M_{1+}$ dominance.

Truncated multipole expansions of the Legendre coefficients used in 
quadrupole estimators are given in Eq. (\ref{eq:Leg-mp}) where the
contributions that satisfy $M_{1+}$ dominance are listed first and where
the remaining terms include some of the lowest multipolarity contributions
of other types but are not necessarily arranged in order of numerical
importance. 
\begin{subequations}
\label{eq:Leg-mp}
\begin{eqnarray}
A_0^L &=& |S_{0+}|^2 + 8 |S_{1+}|^2 + |S_{1-}|^2 + 8 |S_{2-}|^2
+ 27 |S_{2+}|^2 + ... \\
A_0^T &=& 2 |M_{1+}|^2 + |E_{0+}|^2 + |M_{1-}|^2 + 6 |E_{1+}|^2 ] 
+ 6 |M_{2-}|^2 + 2 |E_{2-}|^2 + 9 |M_{2+}|^2 + 18 |E_{2+}|^2 + ... \\
A_0^{TT} &=& -\frac{3}{2} |M_{1+}|^2
-\Re[M_{1+}^* (3 E_{1+} + 3 M_{1-}
+ 12 M_{3-} + 3 E_{3-} + 2 M_{3+} + 10 E_{3+})] \\ \nonumber
&+&  \frac{9}{2}|E_{1+}|^2 + \frac{3}{2}|E_{2-}|^2 + 24 |E_{2+}|^2 
- \frac{9}{2}|M_{2-}|^2 - 12 |M_{2+}|^2 \\ \nonumber
&+& \Re[  -3 E_{0+}^*(E_{2-} + M_{2-} - M_{2+} + E_{2+} ) 
+ E_{1+}^* (3 M_{1-} - 21 E_{3-} - 12 M_{3-} + 12 M_{3+}) \\ \nonumber
&+& M_{1-}^* (3 E_{3-} + 3 M_{3-} + 10 E_{3+} - 10 M_{3+}) + ...] \\
A_1^{LT} &=& 3 \Re[ M_{1+}^* (2 S_{1+} - 3 S_{3-} + 4 S_{3+}) 
+ S_{0+}^* (E_{2-} - M_{2-} + M_{2+} - 4 E_{2+})
 \\ \nonumber
&-& E_{0+}^* (2 S_{2-} - 3 S_{2+} )
-2 E_{1+}^* (S_{1+} + S_{1-}) - 2 M_{1-}S_{1+}^* + ... ] \\
A_2^L &=& 8 |S_{1+}|^2 + 8 |S_{2-}|^2 + \frac{216}{7}|S_{2+}|^2 
+ \Re[S_{0+}^* (8 S_{2-} + 18 S_{2+}) +  8 S_{1+} S_{1-}^* + ...]
\\
A_2^T &=& -|M_{1+}|^2 + \Re[ M_{1+}^*( 6 E_{1+} -2 M_{1-} + \frac{24}{7} M_{3-}
+ 6 E_{3-} + \frac{144}{7} M_{3+}) ] \\ \nonumber
&+& 3|E_{1+}|^2 - |E_{2-}|^2 + \frac{108}{7}|E_{2+}|^2 + 3 |M_{2-}|^2
+ \frac{36}{7} |M_{2+}|^2 \\ \nonumber
&+& \Re[ E_{0+}^*(2 E_{2-} - 6 M_{2-} + 6 M_{2+}+ 12 E_{2+} )
 -6 M_{1-}E_{1+}^* + ...]
\end{eqnarray}
\end{subequations}
Table \ref{table:leg-mp} shows that the number of independent terms in the
multipole expansions of these Legendre coefficients increases very rapidly
with the maximum $\ell$ permitted.
Complete expansions for $\ell_\text{max} \leq 6$ can be found in
Ref. \cite{e91011_mpquad} but as $\ell_\text{max}$ increases they quickly 
become too unwieldy to display here or to use in practical applications.
The Legendre coefficients are usually obtained by numerical integration 
of response functions against Legendre functions instead of by these
algebraic formulas, but both methods do agree.

\begin{table}
\caption{Complexity of multipole expansions of Legendre coefficients.
For each Legendre coefficient used for quadrupole estimators we show
the number of independent terms of the form $\Re(a b^*)$, where $a$ and
$b$ are multipole amplitudes with $\ell \leq \ell_\text{max}$. }
\label{table:leg-mp}
\begin{ruledtabular}
\begin{tabular}{rrrrrrr|rrrrrr}
& \multicolumn{6}{c}{complete} & \multicolumn{6}{c}{$M_{1+}$ dominance} \\
$\ell_\text{max}$ & 
$A_0^L$ & $A_0^T$ & $A_1^{LT}$ & $A_0^{TT}$ & $A_2^L$ & $A_2^T$ & 
$A_0^L$ & $A_0^T$ & $A_1^{LT}$ & $A_0^{TT}$ & $A_2^L$ & $A_2^T$ \\ \hline
1 &  3 &  4 &   8 &   9 &  5 &  9 & 0 & 2 & 2 &  4 & 0 & 4 \\
2 &  5 &  8 &  26 &  27 & 11 & 26 & 0 & 2 & 2 &  4 & 0 & 4 \\
3 &  7 & 12 &  50 &  52 & 17 & 46 & 0 & 2 & 4 &  8 & 0 & 4 \\
4 &  9 & 16 &  84 &  85 & 23 & 66 & 0 & 2 & 4 &  8 & 0 & 6 \\
5 & 11 & 20 & 113 & 116 & 27 & 82 & 0 & 2 & 5 & 11 & 0 & 6 \\
6 & 13 & 14 & 150 & 153 & 31 & 98 & 0 & 2 & 5 & 11 & 0 & 6
\end{tabular}
\end{ruledtabular}
\end{table}

The assumption of $M_{1+}$ dominance omits any terms that do not involve
$M_{1+}$, which strongly inhibits the proliferation of terms but is not
sufficient in itself to extract quadrupole ratios from cross section data.
Combined with $sp$ truncation, these expansions reduce to
\begin{subequations}
\label{eq:Leg-trad}
\begin{eqnarray}
A_0^L &\approx& 0 \\
A_0^T &\approx& 2 |M_{1+}|^2 \\
A_0^{TT} &\approx&  -\Re[(\frac{3}{2}M_{1+} + 3 E_{1+} + 3 M_{1-})M_{1+}^*] \\
A_1^{LT} &\approx& 6 \Re[S_{1+}  M_{1+}^* ] \\ 
A_2^L &\approx& 0 \\
A_2^T &\approx& \Re[( -M_{1+} + 6 E_{1+} -2 M_{1-})M_{1+}^*]
\end{eqnarray}
\end{subequations}
Thus, one obtains the traditional quadrupole estimators
\begin{subequations}
\label{eq:estimators}
\begin{eqnarray}
\tilde{R}_{EM} &=& \frac{3(A_2^T + \epsilon A_2^L)-2 A_0^{TT}}
{12 (A_0^T + \epsilon A_0^L)} \\
\tilde{R}_{SM} &=& \frac{A_1^{LT}}{3(A_0^T + \epsilon A_0^L)}
\end{eqnarray}
\end{subequations}
where $A^L_n$ is included because most experiments have not used
Rosenbluth separation to isolate $A^T_n$.
When Rosenbluth separation is available, one can simply use 
$\epsilon \rightarrow 0$ in Eq. (\ref{eq:estimators}).
Therefore,  it is useful to define the accuracy parameters
\begin{subequations}
\label{eq:superratio}
\begin{eqnarray}
f_{EM}(\ell_\text{max},\epsilon) &=& 
\frac{1}{R_{EM}}
\left( \frac{3(A_2^T + \epsilon A_2^L)-2 A_0^{TT}}
{12 (A_0^T + \epsilon A_0^L)} \right)_{\ell \leq \ell_\text{max}} 
\simeq \frac{\tilde{R}_{EM}}{R_{EM}} \\
f_{SM}(\ell_\text{max},\epsilon) &=& 
\frac{1}{R_{SM}}
\left( \frac{A_1^{LT}}{3(A_0^T + \epsilon A_0^L)}
\right)_{\ell \leq \ell_\text{max}} 
\simeq \frac{\tilde{R}_{SM}}{R_{SM}}
\end{eqnarray}
\end{subequations}
where asymptotic equality refers to the limit 
$\ell_\text{max} \rightarrow \infty$.
Despite their appealing simplicity, it is clear that many contributions 
are omitted and the accuracy of the traditional estimators is a 
numerical issue that can be addressed either theoretically using model 
calculations or experimentally using additional polarization measurements
to extract complex multipole amplitudes directly.

The convergence of these expansions is evaluated in 
Table \ref{table:leg-conv} using $p \pi^0$ multipole amplitudes
for $W = 1.232$ GeV and $Q^2 = 1.0$ (GeV/$c$)$^2$ from MAID2003 
\cite{Drechsel99,MAID2003}.
First, we observe that $A_n^L$ contributions are not negligible: 
the contribution of $A_0^L$ to the denominators of Eq. (\ref{eq:estimators}) 
reduces the estimated quadrupole ratios by about $4\%$ without Rosenbluth
separation when $\epsilon \rightarrow 1$.
(Note that $\epsilon = 0.949$ at $W=1.232$ GeV in Ref. \cite{Kelly05c}.)
Even though $A_2^L$ is much smaller, its effect upon $f_{EM}$ is 
even larger because the strong cancellation between $A_0^{TT}$ 
and $A_2^T + \epsilon A_2^L$ amplifies the dependence on $\epsilon$.
Therefore, the assumption of $M_{1+}$ dominance is not sufficiently
accurate to measure $R_{EM}$ without Rosenbluth separation.
Even with Rosenbluth separation, one should not expect better than
about $15\%$ accuracy for either quadrupole ratio using the 
traditional Legendre analysis (see the bottom of last two columns
of Table \ref{table:leg-conv}). 
Second, it is clear that $sp$ truncation is not valid either
because contributions with $\ell > 1$ are not negligible.
Cancellation between contributions to the numerator of $\tilde{R}_{EM}$
also amplifies truncation errors and convergence is not necessarily
monotonic as $\ell_\text{max}$ increases.
Contributions to $A_0^L$ and $A_0^T$ are nonnegative, but the signs for
other Legendre coefficients are mixed.
While the magnitudes of multipole amplitudes for $\ell>1$ do tend
to decrease, their coefficients in Eq. (\ref{eq:Leg-mp}) tend to 
increase with $\ell$.
Thus, convergence becomes a delicate numerical issue.

Under the present conditions, we find that $\Re(M_{1-}E_{1+}^*)$ is
the most important contribution to $\tilde{R}_{EM}$ neglected by
$M_{1+}$ dominance and is approximately $-40\%$ of the leading term.
Thus, $M_{1+}$ dominance is not very accurate.
The fact that $f_{EM}$ approaches $0.88$ for $\epsilon = 0$ is 
actually nothing more than a lucky conspiracy among the magnitudes 
and signs for a very large number of smaller terms, 
many of which are not especially small individually.
However, most experiments omit Rosenbluth separation.
Similarly, the second most important contribution to $\tilde{R}_{SM}$ 
is $\Re(S_{0+}E_{2-}^*)$ but is only about $6\%$ of the leading term;
hence, $f_{SM}$ converges more rapidly.
The details of this analysis are obviously model dependent, but 
qualitatively similar results are obtained for other models as well.
Although $f_{EM}$ for $\epsilon=0$ is slightly closer to unity than 
$f_{SM}$ for the present analysis, the greater susceptibility of
$\tilde{R}_{EM}$ to truncation errors through its reliance upon 
delicate cancellations suggests that the traditional Legendre 
analysis is intrinsically less reliable for $R_{EM}$ than for $R_{SM}$, 
with or without Rosenbluth separation.

\begin{table}
\caption{Convergence of multipole expansions of Legendre coefficients
and quadrupole estimators.
Multipole amplitudes from MAID2003 for the $p \pi^0$ channel were
used for $W=1.232$ GeV and $Q^2 = 1.0$ (GeV/$c$)$^2$.  
Legendre coefficients are in units of $(\mu b)^{1/2}$.}
\label{table:leg-conv}
\begin{ruledtabular}
\begin{tabular}{rrrrrrr|rrrr}
& & & & & & & \multicolumn{2}{c}{$\epsilon = 0.95$} 
& \multicolumn{2}{c}{$\epsilon = 0$} \\
$\ell_\text{max}$ & 
$A_0^L$ & $A_0^T$ & $A_1^{LT}$ & $A_0^{TT}$ & $A_2^L$ & $A_2^T$ & 
$f_{EM}$ & $f_{SM}$ & $f_{EM}$ & $f_{SM}$ \\ \hline
1 & 0.3339 & 7.599 & -1.483 & -5.220 & 0.141 & -3.769 & 0.300 & 0.938 & 0.581 & 0.977 \\
2 & 0.3377 & 7.624 & -1.395 & -5.156 & 0.102 & -3.916 & 0.736 & 0.879 & 0.960 & 0.916 \\
3 & 0.3384 & 7.628 & -1.323 & -5.137 & 0.079 & -3.871 & 0.714 & 0.833 & 0.894 & 0.868 \\
4 & 0.3384 & 7.628 & -1.315 & -5.130 & 0.081 & -3.859 & 0.696 & 0.828 & 0.878 & 0.863 \\
5 & 0.3384 & 7.628 & -1.308 & -5.130 & 0.081 & -3.857 & 0.693 & 0.824 & 0.876 & 0.859 \\
\end{tabular}
\end{ruledtabular}
\end{table}

It is often argued that the traditional Legendre analysis should
be more accurate for the isospin-3/2 channel than for the $p\pi^0$
reaction because the resonant multipoles should share a common phase
and become pure imaginary at the physical mass, thereby suppressing
background contributions.  
Leaving aside the propagation of errors involved in extracting 
isospin-3/2 amplitudes by combining two independent experiments, 
we can address the intrinsic accuracy of this analysis method 
using model calculations also.
The convergence of the accuracy parameters for isospin-3/2
quadrupole ratios is examined in Table \ref{table:iso3/2}.
Again we find that Rosenbluth separation is required for
$\tilde{R}_{EM}$.
Interestingly, $f_{EM}$ deteriorates as $\ell_\text{max}$ 
increases and the final accuracy of $\tilde{R}_{EM}$ is
worse for isospin-3/2 than for $p\pi^0$ even with $\epsilon=0$.
The cancellations are severe, the method is unstable, and 
calculations for $\tilde{R}_{EM}$ are highly model-dependent. 

\begin{table}
\caption{Convergence of quadrupole estimators for isospin-3/2.
Multipole amplitudes from MAID2003 were
used for $W=1.232$ GeV and $Q^2 = 1.0$ (GeV/$c$)$^2$.  
Legendre coefficients are in units of $(\mu b)^{1/2}$.}
\label{table:iso3/2}
\begin{ruledtabular}
\begin{tabular}{rrrrr}
 & \multicolumn{2}{c}{$\epsilon = 0.95$} 
 & \multicolumn{2}{c}{$\epsilon = 0$} \\
$\ell_\text{max}$ & 
$f_{EM}$ & $f_{SM}$ & $f_{EM}$ & $f_{SM}$ \\ \hline
1 & 0.717 & 0.971 & 0.993 & 1.007 \\
2 & 0.516 & 0.868 & 0.831 & 0.900 \\
3 & 0.484 & 0.881 & 0.801 & 0.914 \\
4 & 0.452 & 0.871 & 0.767 & 0.903 \\
5 & 0.447 & 0.872 & 0.763 & 0.905 \\
\end{tabular}
\end{ruledtabular}
\end{table}

\begin{figure}
\centering
\includegraphics[width=3in]{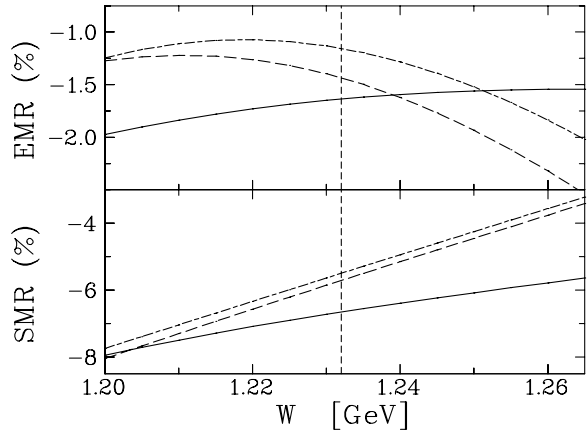}
\caption{$W$ dependence of quadrupole estimators for $p\pi^0$
at $Q^2=1.0$ (GeV/$c$)$^2$ using MAID2003 multipoles.
Solid curves show $R_{EM}$ and $R_{SM}$ while dashed and dash-dotted
curves show $\tilde{R}_{EM}$ and $\tilde{R}_{SM}$ for $\ell \leq 5$
using $\epsilon=0$ and $\epsilon=0.9$, respectively.
The vertical dashed line denotes $W=1.232$ GeV.}
\label{fig:quad_W}
\end{figure}

Figure \ref{fig:quad_W} compares traditional quadrupole estimators 
with $R^{(p\pi^0)}_{EM}$ and $R^{(p\pi^0)}_{SM}$ for MAID2003 at
$Q^2=1.0$ (GeV/$c$)$^2$.
Ideally the estimators would be most accurate in the immediate
vicinity of the physical mass, $W=M_\Delta\approx 1.232$ GeV,
but neither actually has that property.
Rosenbluth separation is most important for $R_{EM}$, but
even with separation the residual error at $M_\Delta$ is significant
at the level of experimental precision that is now possible.

\begin{figure}
\centering
\includegraphics[width=3in]{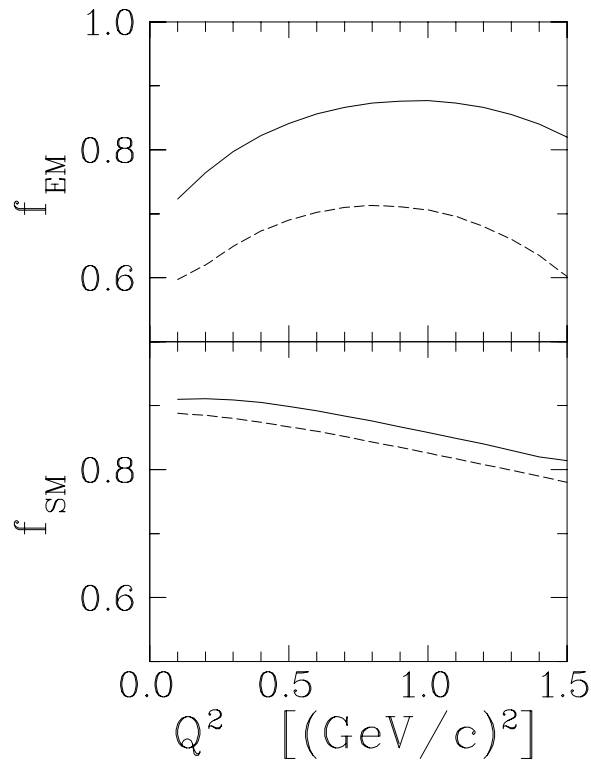}
\caption{Accuracy of traditional quadrupole estimators for $p\pi^0$
at $W=1.232$ GeV using MAID2003 multipoles.
Solid curves use $\epsilon=0$ and dashed curves use $\epsilon=0.9$.}
\label{fig:quad}
\end{figure}

Similarly, Fig.\ \ref{fig:quad} shows the $Q^2$ dependence
for the accuracies of the traditional quadrupole estimators 
using MAID2003 $p\pi^0$ multipole amplitudes 
for $W=1.232$ GeV.
Solid curves use $\epsilon=0$, corresponding to Rosenbluth
separation, while dashed curves use $\epsilon = 0.9$, typical
of many experiments.
Even with Rosenbluth separation, neither quadrupole estimator
can be trusted to better than about $20\%$ and their accuracy deteriorates 
at larger $Q^2$ as $M_{1+}$ dominance breaks down.
Therefore, truncation errors can seriously affect the $Q^2$ dependence 
of quadrupole amplitudes deduced from Legendre coefficients.
The most extensive recent study of $p\pi^0$ quadrupole ratios 
for $0.4 \leq Q^2 \leq 1.8$ (GeV/$c$)$^2$ used Eq. (\ref{eq:estimators}) 
without Rosenbluth separation \cite{Joo02}. 
We do not advocate adjustment of such results using Fig.\ \ref{fig:quad},
at least at this time, because the shapes of $f_{EM}$ and $f_{SM}$ are 
model dependent and MAID2003 does not describe all of the low-lying 
multipole amplitudes at $Q^2 = 1$ (GeV/$c$)$^2$ from Ref. \cite{Kelly05d} 
sufficiently well to be confident of its predictions for the $Q^2$ 
dependencies of these ratios.
Instead, we claim that accurate measurements of the quadrupole ratios 
require multipole analysis of both polarization and cross section data.   

Finally, other simple estimators 
\begin{subequations}
\label{eq:more}
\begin{eqnarray}
\Re E_{0+}M_{1+}^* &\approx& A_1^T / 2  \\
\Re S_{0+}M_{1+}^* &\approx& A_0^{LT} \\
\Re M_{1-}M_{1+}^* &\approx& -(2 A_0^T + 2 A_0^{TT} + A_2^T)/8 
\end{eqnarray}
\end{subequations}
based upon $M_{1+}$ dominance and $sp$ truncation are sometimes
quoted \cite{Joo02,Burkert04}.
Note that Rosenbluth separation is required.  
However, using MAID2003 $p\pi^0$ multipoles at $(W,Q^2)=(1.232,1.0)$ with
$\ell \leq 5$, the ratios between the right- and left-hand sides of
Eq. (\ref{eq:more}) are $1.74$, $-0.77$, and $9.75$.
Most notably, the numerical contribution of $\Re M_{1-}M_{1+}^*$ is 
only the fifth largest term in the multipole expansion of the specified 
combination of Legendre coefficients.
Therefore, these estimators are worthless under these conditions.

In summary, we have performed a detailed numerical analysis of truncation
errors in quadrupole ratios deduced from Legendre coefficients fit to cross 
section angular distributions.
We find that neither $M_{1+}$ dominance nor $sp$ truncation is reliable 
and that one cannot expect better than $20\%$ accuracy from this method.
Truncation errors are especially important for $R_{EM}$.
Furthermore, accurate results for $R_{EM}$ also require Rosenbluth
separation, which was not performed in recent studies of the $Q^2$
dependence of the quadrupole ratios.
The accuracy of truncated Legendre analyses of $E_{0+}$, $S_{0+}$, and
especially $M_{1-}$ is even worse.
Polarization data for pion electroproduction are needed to perform
nearly model-independent multipole analyses that provide complex amplitudes
and do not depend upon unjustifiable truncation schemes.

\begin{acknowledgments}
The support of the U.S. National Science Foundation under grant PHY-0140010
is gratefully acknowledged.
\end{acknowledgments}


\begin{thebibliography}{10}

\bibitem{Blanpied01}
G. Blanpied {\it et~al.}, Phys. Rev. {\bf C} {\bf {\bf 64}},  025203  (2001).

\bibitem{Joo02}
K. Joo {\it et~al.}, Phys. Rev. Lett. {\bf {\bf 88}},  122001  (2002).

\bibitem{Sparveris05}
N.~F. Sparveris {\it et~al.}, Phys. Rev. Lett. {\bf {\bf 94}},  022003  (2005).

\bibitem{Frolov99}
V.~V. Frolov {\it et~al.}, Phys. Rev. Lett. {\bf {\bf 82}},  45  (1999).

\bibitem{Kalleicher97}
F. Kalleicher, U. Dittmayer, R.~W. Gothe, H. Putsche, T. Reichelt, B. Schoch,
  and M. Wilhelm, Zeit. Phys. {\bf A} {\bf {\bf 359}},  201  (1997).

\bibitem{Kelly05c}
J.~J. Kelly {\it et~al.}, arXiv:nucl-ex/0505024 (unpublished).

\bibitem{Kelly05d}
J.~J. Kelly {\it et~al.}, to be submitted to PRC (unpublished).

\bibitem{e91011_mpquad}
J.~J. Kelly, {\it Mathematica} notebook at
  \url{www.physics.umd.edu/enp/jjkelly/e91011/Reports/Notebooks/QuadrupoleRati%
os.nb} (unpublished).

\bibitem{Drechsel99}
D. Drechsel, O. Hanstein, S.~S. Kamalov, and L. Tiator, Nucl. Phys. {\bf {\bf
  A645}},  145  (1999).

\bibitem{MAID2003}
D. Drechsel, O. Hanstein, S.~S. Kamalov, L. Tiator, and S.~N. Yang,
  \url{www.kph.uni-mainz.de/maid/maid2003}.

\bibitem{Burkert04}
V.~D. Burkert and T.-S.~H. Lee, Int. Journ. Mod. Phys. {\bf E} {\bf {\bf 13}},
  1035  (2004).

\end{thebibliography}

\end{document}